\documentclass{article}
\usepackage[margin=2.5cm, includefoot, footskip=30pt]{geometry}

\usepackage{amsmath}
\usepackage{booktabs}
\usepackage{graphics}
\usepackage{multicol}
\usepackage[ruled,vlined]{algorithm2e}
\usepackage{setspace}
\usepackage{graphicx}
\usepackage{subcaption}
\usepackage{hyperref}
\usepackage{color,colortbl}
\usepackage{array}
\usepackage{booktabs}
\usepackage{tabularx}
\usepackage{wrapfig, blindtext}
\usepackage{soul}
\usepackage{enumerate}
\usepackage[table]{xcolor}
\usepackage{authblk}
\usepackage{pdfpages}

\definecolor{Gray}{gray}{0.92}
\usepackage[first=0,last=9]{lcg}

\def\axelrod{\texttt{Axelrod-Python}}
\def\TFT{\texttt{TFT}}
\def\IPD{\texttt{IPD}}

\title{Properties of Winning Iterated Prisoner's Dilemma Strategies.}
\author[1, 2]{Nikoleta E. Glynatsi}
\author[2, $\dagger$]{Vincent Knight}
\author[3, $\dagger$]{Marc Harper}

\affil[1]{Max Planck Institute for Evolutionary Biology, Research Group Dynamics of Social
Behavior, Germany}
\affil[2]{Cardiff University, School of Mathematics, UK}
\affil[3]{Google Inc., Mountain View, CA, USA}
\affil[$\dagger$]{V.K. and M.H. contributed equally to this work.}
\date{}

\begin{document}

\maketitle

\begin{abstract}
Researchers have explored the performance of Iterated Prisoner's Dilemma strategies
for decades, from the celebrated performance of Tit for Tat to the
introduction of the zero-determinant strategies and the use of sophisticated learning
structures such as neural networks. Many new strategies have been introduced and tested
in a variety of tournaments and population dynamics. Typical results in the literature,
however, rely on performance against a small number of somewhat arbitrarily selected
strategies in a small number of tournaments, casting doubt on the generalizability
of conclusions. In this work, we analyze a large collection of 195
strategies in thousands of computer tournaments, present the top performing strategies across multiple
tournament types, and distill their salient features.
The results show that there is not yet a single
strategy that performs well in diverse Iterated Prisoner's Dilemma scenarios,
nevertheless there are several properties that heavily influence the best performing
strategies. This refines the properties described by Axelrod in light of
recent and more diverse opponent populations to: be nice, be provocable and generous,
be a little envious, be clever, and adapt to the environment. More precisely,
we find that strategies perform best when their probability of cooperation
matches the total tournament population's aggregate cooperation probabilities.
The features of high performing strategies help cast some light on why
strategies such as Tit For Tat performed historically well in tournaments and
why zero-determinant strategies typically do not fare well in tournament
settings. Furthermore, our findings have implications for the future training of
autonomous agents, as understanding the crucial features for incorporation into
these agents becomes essential.
\end{abstract}

\section{Background}

The Iterated Prisoner's Dilemma (\IPD) is a repeated two-player game that models
behavioral interactions, specifically interactions where self-interest clashes
with collective interest. In each turn of the game, both players simultaneously
and independently decide between cooperation (\(C\)) and defection (\(D\)). This
decision is made with the memory of all prior interactions. The payoffs for each
player at each turn are influenced by their own choice and the choice of the
other player. To this end, the payoffs of the game are defined by

\begin{center}
{\renewcommand{\arraystretch}{2}%
\begin{tabular}{c|c|c}
& Cooperate (\(C\)) & Defect (\(D\)) \\
\hline
Cooperate (\(C\)) & \(R, R\) & \(S, T\) \\
\hline
Defect (\(D\)) & \(T, S\) & \(P, P\) \\
\end{tabular}}
\end{center}

where typically \(T > R > P > S\) and \(2R > T + S\). The most common values
used in the literature~\cite{Axelrod1981} are \(R=3\), \(P=1\), \(T=5\),
\(S=0\), and these are the values also used in this work.

Conceptualizing strategies and understanding the best way to play the game have
been of interest to the scientific community since the formulation of the game
in 1950~\cite{Flood1958}. Following Axelrod's computer tournaments in the
1980s~\cite{Axelrod1980a, Axelrod1980b}, round-robin computer tournaments became
a common evaluation technique for newly designed strategies.
The winner of both of Axelrod's tournaments~\cite{Axelrod1980a, Axelrod1980b}
was the simple strategy Tit For Tat (\TFT{}). \TFT{} cooperates on the first turn
and thereafter copies the previous action of its opponent, retaliating against
defections with a defection and forgiving a defection if followed by
cooperation. Axelrod concluded that the strategy's robustness was due to four
properties, which he adapted into four suggestions for success in an \IPD{}
tournament:

\begin{enumerate}[(i)]
    \item Do not be envious by striving for a payoff larger than the opponent's payoff.
    \item Be ``nice'' by not being the first to defect.
    \item Reciprocate both cooperation and defection; Be provocable to retaliation and forgiveness.
    \item Do not be too clever by scheming to exploit the opponent.
\end{enumerate}

Forgiveness, in this context, is a strategy's ability to cooperate after a
\(DC\) outcome to achieve mutual cooperation again. In environments without
noise, \TFT{} would end up in \(DC\) only if it had received a defection and
then retaliated. Subsequently, \TFT{} would forgive an opponent that apologizes
(in a \(DC\) round) by returning to cooperation, as mutual cooperation is deemed
better than mutual defection.

Due to the strategy's strong performance in both tournaments and a series of
evolutionary experiments~\cite{Axelrod1981}, \TFT{} was often claimed to be a
highly robust (and sometimes the most robust) strategy for the \IPD.
There are strategies that have built upon \TFT{} and the reciprocity-based
approach. In~\cite{Beaufils1997}, a strategy called Gradual was introduced,
constructed to have the same qualities as those of \TFT{} with one
addition. Gradual has a memory of the previous rounds of play in the game,
recording the number of defections by the opponent and punishing them with a
growing number of defections. It then enters a calming state in which it
cooperates for two rounds. A strategy with the same intuition as Gradual is
Adaptive Tit for Tat~\cite{tzafestas-2000a}. Adaptive Tit for Tat maintains a
continually updated estimate of the opponent's behavior and uses this estimate
to condition its future actions.
Other research has built upon the limitations of \TFT. For example,
in~\cite{Bendor1991, Donninger1986, Molander1985, Hammerstein1984}, it was shown
that \TFT{} suffered in environments with noise. This was mainly due to the
strategy being too provocable and its lack of generosity and contrition. Since
\TFT{} immediately punishes a defection, in a noisy environment, it can get
stuck in a repeated cycle of defections and cooperations. Some new strategies,
more robust in tournaments with noise, were soon introduced, including Nice and
Forgiving~\cite{Bendor1991}, Generous Tit For Tat~\cite{Nowak1992}, and Pavlov
(aka Win Stay Lose Shift)~\cite{Nowak1993}, as well as later variants such as
OmegaTFT \cite{kendall2007iterated}.

Finally, others introduced strategies deviating completely from the originally
suggested properties of success. For example, a set of ``envious'' Iterated
Prisoner's Dilemma (\IPD) strategies were introduced, called zero-determinant
strategies (ZDs), in~\cite{Press2012}. These strategies attempt to force a
linear relationship between stationary payoffs against other memory-one
opponents, potentially ensuring that they receive a higher average payout. While
ZDs were introduced with a small tournament in which some were reportedly
successful \cite{Stewart2012}, this result has not generally held in future
work~\cite{mathieu2017}. Furthermore, in~\cite{Harper2017}, a series of ``clever''
strategies trained using reinforcement learning were introduced. These
strategies were trained using lookup tables~\cite{Axelrod1987}, hidden Markov
models~\cite{Harper2017}, and finite-state automata~\cite{Miller1996}, on
a set of 170 strategies.

One thing that has remained the same is that the introduction of a new strategy
is often accompanied by a claim that the new strategy is the best performing
strategy for the \IPD, often without extensive testing against a broad spectrum
of opponents or representative classes of opponents. The lack of testing against
formally defined strategies and tournament winners is understandable given the
effort required to implement the hundreds of published \IPD{} strategies.
Implementing prior strategies faithfully is often extremely difficult or
impossible due to insufficient descriptions and lack of published
implementations or code. Despite these challenges, the absence of thorough
testing raises concerns about claims regarding the superiority or robustness of
newly introduced strategies.

In this paper, we evaluate the performance of a significant number of \IPD{}
strategies across a diverse array of tournaments. Many of the strategies used in
our analysis are drawn from well-known and named strategies in \IPD{}
literature, including previous tournament winners. This contrasts with other
work that may have randomly generated essentially arbitrary strategies, often
constrained to specific classes such as memory-one strategies or those of a
certain structural form like finite state machines or deterministic memory-two
strategies. Furthermore, our tournaments encompass variations, including standard
tournaments resembling Axelrod's original ones, tournaments with noise,
probabilistic match length, and both noise and probabilistic match length. This
diversity in strategies and tournament types provides new insights and tests
earlier claims in alternative settings against known powerful strategies.
More specifically, we show that the previous tournament
winners are lacking against large enough opponent pools; they do not appear
among the top-performing strategies anymore. This could be due to likely
suffering from a lack of diversity in the strategies they were trained/tested
against, finding it hard to adapt to the new strategies.

It is important to note that we do not assert the existence of a single
best-performing strategy across all tournaments or tournament types. On the
contrary, our work demonstrates that such a strategy does not exist
(notwithstanding a few strategies with broadly high performance). The primary
objective of this paper, presented in the latter parts of the paper, is to
continue the discussion on the properties of successful strategies, a
conversation started by Axelrod. 
The results of our analysis conclude that the properties of a successful
strategy in the Iterated Prisoner's Dilemma (\IPD{}) are:

\begin{enumerate}[(i)]
    \item Be a little bit envious
    \item Be ``nice'' in non-noisy environments or when game lengths are longer
    \item Reciprocate both cooperation and defection appropriately;
    Be provocable in tournaments with short matches, and generous in tournaments with noise
    \item It's ok to be clever
    \item Adapt to the environment; Adjust to the mean population cooperation
\end{enumerate}

We believe that the discussion on the properties of winning strategies holds
significant importance. It aims to provide guidance to researchers designing new
strategies and those training strategies. Specifically, much like the recognized
value of diversity in training datasets, such as variations in image perspective,
skin color, etc., are critical in training accurate and generalizable machine
learning models, we show that diversity in the population of opponent strategies
is of paramount importance in the construction and evaluation of game theory
strategies. As AI agents start to interact as scale, they will benefit from
exposure to a wide variety of alternative agents. Moreover, conducting a similar
analysis can shed light on already trained strategies, aiding in understanding
the key features they have autonomously developed during their training
processes.

The rest of the paper is organized as follows. In
section~\ref{section:data_collection}, we describe the data collection process.
In subsection~\ref{section:top_performances}, we present the best performing
strategies for each type of tournament.
Subsection~\ref{section:evaluation_of_performance} explores the traits that
contribute to good performance, and in
subsection~\ref{section:winning_features}, we focus on the features of the
winners of the tournaments. Finally, the results are summarized in
section~\ref{section:conclusion}. This manuscript introduces several parameters,
which are discussed in the following sections; the full set of parameters and
their definitions are also provided in the Supplementary Material.

\section{Data collection}\label{section:data_collection}

The data collection of various types of tournaments and the use of different
strategies are made possible due to an open-source library called \axelrod{}~\cite{axelrodproject}
(version 3.0.0). \axelrod{} enables the simulation of \IPD{}
tournaments and contains an extensive list of strategies. Most of these
strategies are described in the literature, with a few exceptions contributed
specifically to the package. In this paper, we use a total of
195 strategies, which can be found in the Supplementary
Material. The package supports several tournament types, and this work considers
standard, noisy, probabilistic ending, and noisy probabilistic ending
tournaments.

{\it Standard tournaments} are similar to Axelrod's well-known
tournaments~\cite{Axelrod1980a}. In these tournaments, there are \(N\)
strategies, and each strategy plays an iterated game with \(n\) turns against
all other strategies, not including self-interactions.
{\it Noisy tournaments} also involve \(N\) strategies and \(n\) turns, but
in each turn, there is a probability \(p_n\) that a player's action is
flipped. Compared to these two tournaments, in 
{\it probabilistic ending tournaments} the number of turns is not fixed. Instead,
a match between strategies ends with a given probability
\(p_e\).
Finally, {\it noisy probabilistic ending tournaments} incorporate both a
noise probability \(p_n\) and an ending probability \(p_e\). For smoother
results, each tournament is repeated \(k\) times, and this repetition factor was
allowed to vary to assess the impact of smoothing. The winner of each tournament
is determined based on the average score achieved by a strategy from the entire
set of repetitions, not by the number of wins.

The process of collecting tournament results is outlined in
Algorithm~\ref{algorithm:data_generation}. For each trial, a random size \(N\)
is selected, and a random list of \(N\) strategies from the 195
available. Subsequently, one standard, one noisy, one probabilistic ending, and
one noisy probabilistic ending tournament are conducted for the selected list of
strategies. The parameters for the tournaments, as well as the number of
repetitions, are chosen once for each trial. We have run a total of
11400 trials of Algorithm~\ref{algorithm:data_generation}. For
each trial, we collect the results for four different tournaments, resulting in
a total of 45606 $(11400 \times 4)$ tournament
results. Each tournament outputs a result summary in the form of
Table~\ref{table:output_result}.

\begin{algorithm}[!htbp]
    \setstretch{1.35}
    \For{\text{seed} $\in [0, 11420]$}{
        $N \gets \text{randomly select integer}\in [3, 195]$\;
        $\text{players} \gets  \text{randomly select $N$ players}$\;
        $k \gets  \text{randomly select integer}\in [10, 100]$\;
        $n \gets  \text{randomly select integer}\in [1, 200]$\;
        $p_n \gets  \text{randomly select float}\in [0, 1]$\;
        $p_e \gets   \text{randomly select float}\in [0, 1]$\;
        \vspace{0.4cm}
        $\text{result standard}$ $\gets$ Axelrod.tournament$(\text{players}, n, k)$\;
        $\text{result noisy}$ $\gets$ Axelrod.tournament$(\text{players}, n, p_n, k)$\;
        $\text{result probabilistic ending}$ $\gets$ Axelrod.tournament$(\text{players}, p_e, k)$\;
        $\text{result noisy probabilistic ending}$ $\gets$ Axelrod.tournament$(\text{players}, p_n, p_e, k)$\;

    }
    \KwRet{result standard, result noisy, result probabilistic ending,
    result noisy probabilistic ending}\;
    \caption{Tournament Data Collection Algorithm}
    \label{algorithm:data_generation}
\end{algorithm}

\newcolumntype{g}{>{\columncolor{Gray}}c}
\begin{table}[b]
    \begin{center}
    \resizebox{\textwidth}{!}{
    \begin{tabular}{ccccccgcgcgcgcg}
    \toprule
    & & & & & &   \multicolumn{8}{g}{Rates}  \\
    Rank & Name & Median score & Cooperation rating $(C_r)$ & Win & Initial C &
    CC & CD & DC & DD & CC to C & CD to C & DC to C & DD to C \\
    0 &  EvolvedLookerUp2 2 2 & 2.97 & 0.705 & 28.0 & 1.0 & 0.639 & 0.066 & 0.189 &
    0.106 & 0.836 & 0.481 & 0.568 & 0.8 \\
    1 &  Evolved FSM 16 Noise 05 & 2.875 & 0.697 & 21.0 & 1.0 & 0.676 &
    0.020 & 0.135 & 0.168 & 0.985 & 0.571 & 0.392 & 0.07 \\
    2 & PSO Gambler 1 1 1 & 2.874 & 0.684 &  23.0 &     1.0 &    0.651 &    0.034 &    0.152 &    0.164
    & 1.000 & 0.283 & 0.000 & 0.136 \\
    3 &  PSO Gambler Mem1 &  2.861 &        0.706 &  23.0 &      1.0 &    0.663
    &    0.042 &    0.145 &    0.150 &  1.000 &  0.510 &  0.000 &  0.122 \\
    4 &          Winner12 &  2.835 &        0.682 &  20.0 &      1.0 &
    0.651 &    0.031 &    0.141 &    0.177 &  1.000 &  0.441 &  0.000 &  0.462 \\
    $\dots$ & $\dots$ & $\dots$ & $\dots$ & $\dots$ & $\dots$ & $\dots$ & $\dots$ &
    $\dots$ & $\dots$ & $\dots$ & $\dots$ & $\dots$ & $\dots$ \\
    \bottomrule
    \end{tabular}}
\end{center}
\caption{\textbf{Result Summary Example of a Tournament.}
A result summary consists of \(N\) rows, with each row containing information
for each strategy that participated in the tournament. This information includes
the strategy's rank (\(R\)), median score, the cooperation rate (\(C_r\)), the number of
match wins, and the probability that the strategy cooperated in the opening
move. Additionally, it provides the probabilities of a strategy being in any of
the four states ($CC, CD, DC, DD$) and the cooperation rate after each state.}\label{table:output_result}
\end{table}

During the data collection process, the probabilities of noise (\(p_n\)) and
tournament ending (\(p_e\)) were allowed to take values between 0 and 1.
However, commonly used values for these probabilities are \(p_n \leq 0.1\) and
\(p_e \leq 0.1\). This is to make the results more interpretable. For example,
consider a strategy competing in an environment with \(p_n > 0.1\). In cases
with a high value of noise, most of the actions the strategy takes are the
complete opposite of what the strategy is designed to do. Therefore, we will
focus on the tournaments for which \(p_n \leq 0.1\) and \(p_e \leq 0.1\). Thus,
the results presented here pertain to subsets of the noisy and probabilistic
ending tournaments. Specifically, the results rely on 1150 tournaments with
noise, 1134 tournaments with a probabilistic ending, and 117 tournaments with
both noise and a probabilistic ending. We also provide an analysis of the paper
considering the entire datasets, and these results are presented in the
Supplementary Material. The general results of the analysis are not affected by
the restriction of the noise and probabilistic ending probabilities.

\section{Results}
\subsection{Top ranked strategies across tournaments}\label{section:top_performances}

A strategy has participated in multiple tournaments of each type, and to
evaluate its overall performance, we introduce a measure called the {\it
normalized rank}. In each tournament, the strategies receive a rank (\(R\)), where 0
denotes that the strategy was the winner, and \(N-1\) indicates that the
strategy came last in the tournament. The normalized rank, denoted as \(r\), is
calculated as \(r = \frac{R}{N-1}\). Thus, the rank a strategy achieved over the
number of players in the tournament. The performance of the strategies is
assessed based on the {\it median of the normalized rank}, denoted as
\(\bar{r}\).

\begin{figure}[!htbp]
    \centering
    \includegraphics[width=\textwidth]{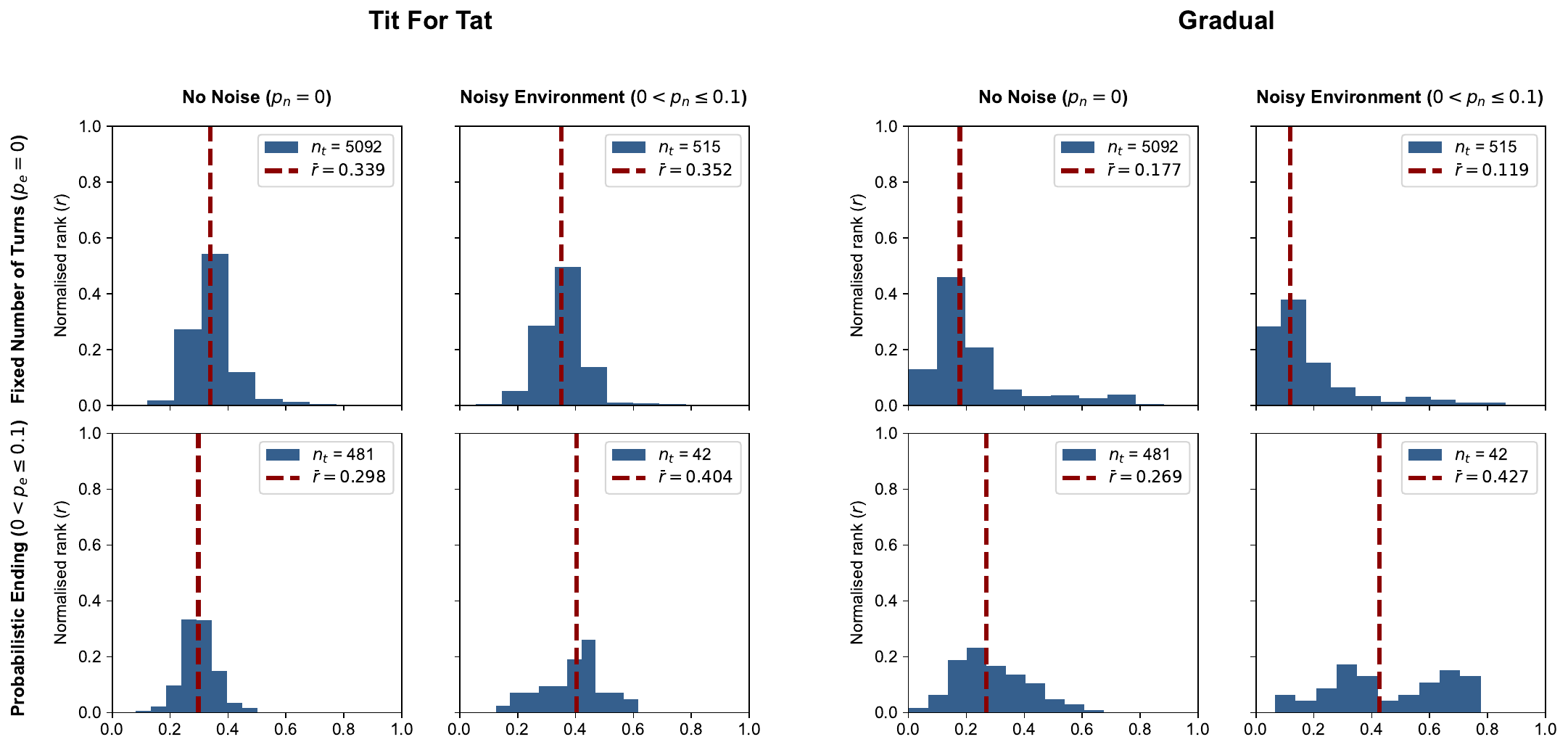}
    \caption{\textbf{Examples of normalized rank distributions for two
    strategies, \TFT{} and Gradual.} We plot the distributions of \(r\) for the
    two strategies in the four tournament types. As a reminder, lower values of
    \(r\) correspond to better performances. The top left quadrant of each plot
    shows the distribution for standard tournaments (fixed number of turns and
    no noise). The top right quadrant shows the distribution for noisy
    tournaments (fixed number of turns and noise). The bottom left quadrant
    shows the distribution for probabilistic ending tournaments (no noise and
    probabilistic ending). Finally, the bottom right quadrant shows the
    distribution for noisy probabilistic ending tournaments (noise and
    probabilistic ending). In each quadrant, we also show the number of data
    points. Both strategies participated in a similar number of tournaments.
    Based on the median rank, which we use in this work to define overall
    performance, \TFT{} performs best in probabilistic ending tournaments,
    whereas Gradual was in standard tournaments.}
    \label{fig:normalised_rank_distributions}
\end{figure}

For example, let's consider the well-known strategies \TFT{} and Gradual. Each
strategy participated in several tournaments of each type. In
Figure~\ref{fig:normalised_rank_distributions} we show the distribution of the
normalised ranks of these strategies in each of the four tournaments. We can
observe that \TFT{} looks to be normally distributed normalized rank. In comparison,
Gradual's performance has longer tails, indicating that there were tournaments
where the strategy performed very well or very poorly. Overall, Gradual achieves
a lower median rank, signifying that it performs better than \TFT{} except in
the case of noisy and probabilistic ending tournaments (lower rank is better).

\begin{figure}[!htbp]
    \centering
    \includegraphics[width=\textwidth]{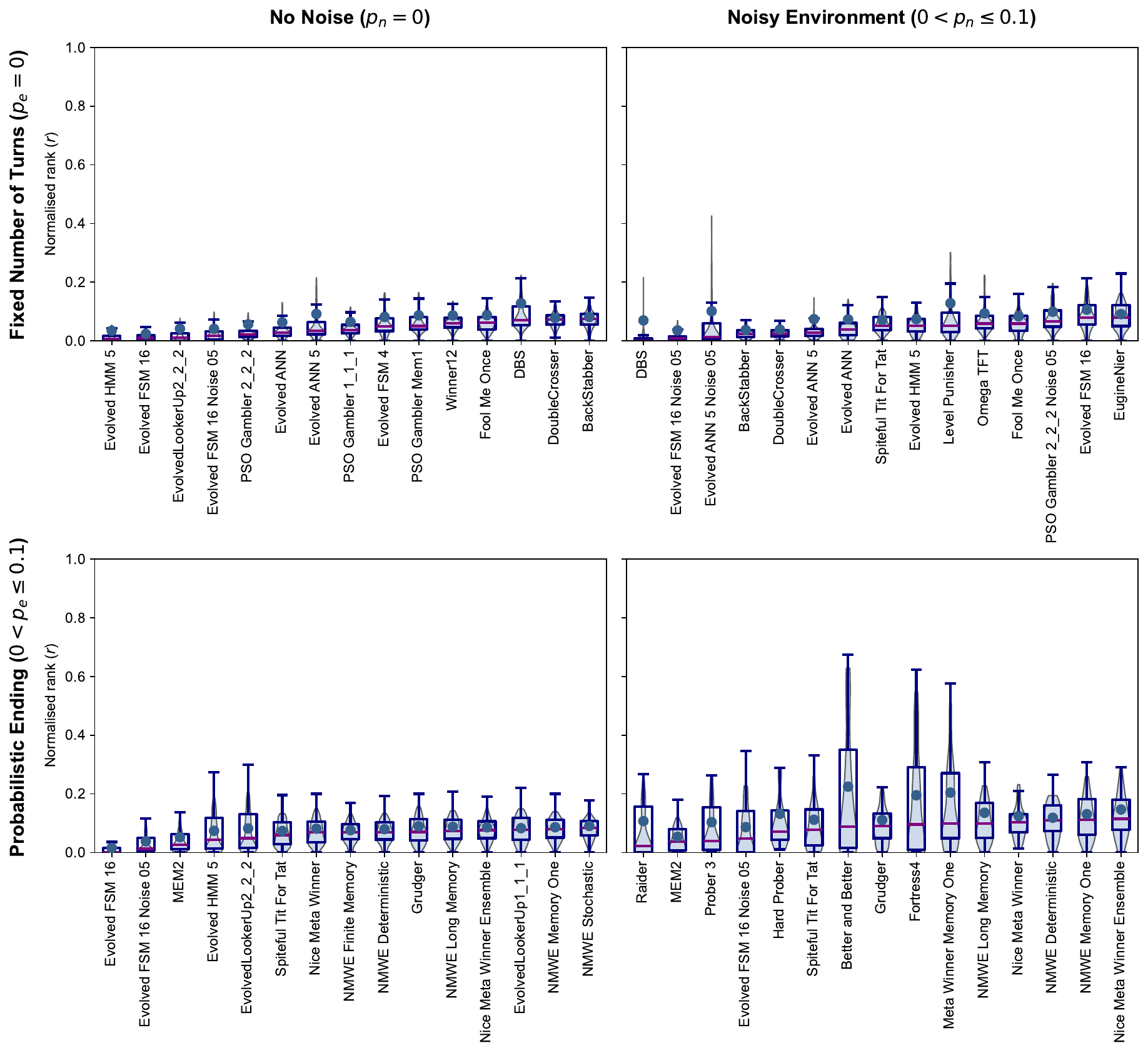}
    \caption{\textbf{\(r\) distributions of the top 15 strategies in different
    environments.} A lower value of \(\bar{r}\) corresponds to a more successful
    performance. A strategy's \(r\) distribution skewed towards zero indicates
    that the strategy ranked highly in most tournaments it participated in. Most
    distributions are skewed towards zero.}\label{fig:r_distributions}
\end{figure}

\newcolumntype{g}{>{\columncolor{Gray}}l}
\begin{table}[!htbp]
    \begin{center}
    \resizebox{\textwidth}{!}{
        
\begin{tabular}{lggllggllggllr}
    \toprule
    & \multicolumn{2}{g}{Standard} & \multicolumn{2}{g}{Noisy (\(p_n \leq 0.1\))} & \multicolumn{2}{g}{Probabilistic ending (\(p_e \leq 0.1\))} &  \multicolumn{2}{c}{Noisy probabilistic ending} \\
    \toprule
    {} &                     Name & $\bar{r}$ &                       Name & $\bar{r}$ &                       Name & $\bar{r}$ &                       Name & $\bar{r}$ \\
    \midrule
    0  &            Evolved HMM 5 &     0.007 &                         DBS &       0.0 &             Evolved FSM 16 &       0.0 &                     Raider &     0.022 \\
    1  &           Evolved FSM 16 &      0.01 &     Evolved FSM 16 Noise 05 &     0.008 &    Evolved FSM 16 Noise 05 &     0.013 &                       MEM2 &     0.037 \\
    2  &     EvolvedLookerUp2\_2\_2 &     0.011 &      Evolved ANN 5 Noise 05 &     0.013 &                       MEM2 &     0.027 &                   Prober 3 &     0.039 \\
    3  &  Evolved FSM 16 Noise 05 &     0.017 &                 BackStabber &     0.024 &              Evolved HMM 5 &     0.043 &    Evolved FSM 16 Noise 05 &     0.048 \\
    4  &        PSO Gambler 2\_2\_2 &     0.022 &               DoubleCrosser &     0.025 &       EvolvedLookerUp2\_2\_2 &     0.049 &                Hard Prober &     0.072 \\
    5  &              Evolved ANN &     0.029 &               Evolved ANN 5 &     0.028 &       Spiteful Tit For Tat &     0.059 &       Spiteful Tit For Tat &     0.078 \\
    6  &            Evolved ANN 5 &     0.034 &                 Evolved ANN &     0.038 &           Nice Meta Winner &     0.069 &          Better and Better &     0.089 \\
    7  &        PSO Gambler 1\_1\_1 &     0.037 &        Spiteful Tit For Tat &     0.051 &         NMWE Finite Memory &     0.069 &                    Grudger &     0.091 \\
    8  &            Evolved FSM 4 &     0.049 &               Evolved HMM 5 &     0.051 &         NMWE Deterministic &      0.07 &                  Fortress4 &     0.096 \\
    9  &         PSO Gambler Mem1 &      0.05 &              Level Punisher &     0.052 &                    Grudger &      0.07 &     Meta Winner Memory One &     0.099 \\
    10 &                 Winner12 &      0.06 &                   Omega TFT &     0.059 &           NMWE Long Memory &     0.074 &           NMWE Long Memory &     0.099 \\
    11 &             Fool Me Once &     0.061 &                Fool Me Once &     0.059 &  Nice Meta Winner Ensemble &     0.076 &           Nice Meta Winner &     0.104 \\
    12 &                      DBS &     0.071 &  PSO Gambler 2\_2\_2 Noise 05 &     0.067 &       EvolvedLookerUp1\_1\_1 &     0.077 &         NMWE Deterministic &     0.109 \\
    13 &            DoubleCrosser &     0.072 &              Evolved FSM 16 &     0.078 &            NMWE Memory One &      0.08 &            NMWE Memory One &     0.112 \\
    14 &              BackStabber &     0.075 &                  EugineNier &      0.08 &            NMWE Stochastic &     0.085 &  Nice Meta Winner Ensemble &     0.115 \\
    \bottomrule
    \end{tabular}
    
    }
\end{center}
\caption{Top performances for each tournament type based on $\bar{r}$. The
results of each type are based on 11420 unique tournaments. The
results for noisy tournaments with \(p_n < 0.1\) are based on 1151 tournaments,
and for probabilistic ending tournaments with \(p_e < 0.1\) on 1139. The top
ranks indicate that trained strategies perform well in a variety of
environments, but so do simple deterministic strategies. For noisy tournaments
DBS is the top ranked strategy with \(\bar{r}=0\), thus DBS won every tournament
it participated in. The same for Evolved FSM 16 Noise 05 in probabilistic ending.}
\label{table:top_performances}
\end{table}

The top 15 strategies for each tournament type, based on \(\bar{r}\), are
presented in Table~\ref{table:top_performances}, while the \(r\) distributions
for the top-ranked strategies can be found in Figure~\ref{fig:r_distributions}.

In standard tournaments dominating strategies were those
trained using reinforcement learning techniques. 10 out of the 15 top strategies
were introduced in~\cite{Harper2017}. These strategies are based on finite state
automata (FSM), hidden Markov models (HMM), artificial neural networks (ANN),
lookup tables (LookerUp), and stochastic lookup tables (Gambler). They have been
trained using reinforcement learning algorithms (evolutionary and particle swarm
algorithms) to perform well against a subset of the strategies in \axelrod{} in a
standard tournament. Thus, their performance in the specific setting was
anticipated, although still noteworthy given the random sampling of tournament
participants. DoubleCrosser and BackStabber, both from the \axelrod, use the
number of turns and are set to defect in the last two rounds. These strategies
can be characterized as “cheaters” because their source code allows them to know
the number of turns (unless the match has a probabilistic ending). These
strategies were expected to not perform as well in tournaments where the number
of turns is not specified. Finally, Winner 12~\cite{mathieu2017} and
DBS~\cite{Au2006} are both from the literature. DBS is a strategy specifically
designed for noisy environments; however, it ranks highly in standard
tournaments as well. Similarly, the fourth-ranked player, Evolved FSM 16 Noise
05, was trained for noisy tournaments yet performs well in standard tournaments.

In the case of noisy tournaments, the
top-performing strategies include strategies specifically designed for noisy
tournaments. These are DBS, Evolved FSM 16 Noise 05, Evolved ANN 5 Noise 05, PSO
Gambler 2 2 2 Noise 05, and Omega Tit For Tat~\cite{kendall2007iterated}. Omega
TFT, a strategy designed to break the deadlocking cycles of \(CD\) and \(DC\)
that TFT can fall into in noisy environments, places 10th. The rest of the top
ranks are occupied by strategies that performed well in standard tournaments and
deterministic strategies such as Spiteful Tit For Tat~\cite{prison}, Level
Punisher~\cite{Eckhart2015}, Eugine Nier~\cite{lesswrong}.

Furthermore, in tournaments with probabilistic endings, the highly ranked
strategies leaned towards defecting strategies and trained finite state
automata, as demonstrated by the works of Ashlock et
al.\cite{Ashlock2006,Ashlock2014}. The most effective strategies in
probabilistic ending tournaments are also a series of ensemble Meta strategies,
trained strategies that performed well in standard tournaments, and
Grudger\cite{axelrodproject} and Spiteful Tit for Tat~\cite{prison}. The Meta
strategies~\cite{axelrodproject} utilize a team of strategies and aggregate the
potential actions of the team members into a single action in various ways.

While no single strategy consistently outperforms all others in any of the
distinct tournament types or across various tournament types, certain types of
strategies consistently achieve top rankings. These include strategies that have
undergone training, those that retaliate, and those that adapt their behavior
based on preassigned rules to optimize outcomes. These findings challenge some
of Axelrod's suggestions, particularly the advice to ``not be clever'' and ``not
be envious''.

\subsection{The effect of strategy features on performance}\label{section:evaluation_of_performance}

For each strategy, we have a variety of features as described in
Table~\ref{table:manual_features}. These features capture measures related to a
strategy's behavior in the tournaments it competed in, as well as intrinsic
properties, such as whether a strategy is deterministic or stochastic. The
correlation coefficients between the features for performance evaluation, the
median score and the median normalised rank are given by
Table~\ref{table:correlations}. The correlation coefficients between all
features have also been calculated and a graphical representation can be found
in the Supplementary Material.

\newcolumntype{g}{>{\columncolor{Gray}}c}
\begin{table}[!htbp]
    \begin{center}
    \resizebox{.99\textwidth}{!}{
    \begin{tabular}{gcgcgc}
    \toprule
    feature & feature explanation &  source & value type & min value & max value \\
    \midrule
stochastic  &  If a strategy is stochastic & strategy classifier from APL & boolean  & Na &  Na \\
makes use of game &  If a strategy makes used of the game information & strategy classifier from APL & boolean  & Na &  Na \\
makes use of length &  If a strategy makes used of the number of turns & strategy classifier from APL & boolean  & Na &  Na \\
memory usage &  The memory size of a strategy divided by the number of turns & memory size from APL & float & 0 &  1 \\
SSE & A measure of how far a strategy is from ZD behaviour & method described in~\cite{Knight2019} & float & 0 & 1 \\
max cooperating rate $(C_{\text{max}})$  & The biggest cooperating rate in a given tournament  & result summary  & float & 0 & 1\\
min cooperating rate $(C_{\text{min}})$ & The smallest cooperating rate in a given tournament  & result summary  & float & 0 & 1\\
median cooperating rate $(C_{\text{median}})$ & The median cooperating rate in a given tournament  & result summary  & float & 0 & 1\\
mean cooperating rate $(C_{\text{mean}})$ & The mean cooperating rate in a given tournament  & result summary  & float & 0 & 1 \\
$C_r$ / $C_{\text{max}}$ & A strategy's cooperating rate divided by the maximum & result summary  & float & 0 & 1 \\
$C_{\text{min}}$ / $C_r$ & A strategy's cooperating rate divided by the minimum & result summary  & float & 0 & 1 \\
$C_r$ / $C_{\text{median}}$ & A strategy's cooperating rate divided by the median  & result summary  & float & 0 & 1\\
$C_r$ / $C_{\text{mean}}$ & A strategy's cooperating rate divided by the mean & result summary  & float & 0 & 1 \\
$C_r$ & The cooperating ratio of a strategy & result summary  & float & 0 & 1 \\
$CC$ to $C$ rate & The probability a strategy will cooperate after a mutual cooperation & result summary  & float & 0 & 1\\
$CD$ to $C$ rate & The probability a strategy will cooperate after being betrayed by the opponent & result summary  & float & 0 & 1 \\
$DC$ to $C$ rate & The probability a strategy will cooperate after betraying the opponent & result summary  & float & 0 & 1 \\
$DD$ to $C$ rate & The probability a strategy will cooperate after a mutual defection & result summary  & float & 0 & 1 \\
$p_n$ & The probability of a player's action being flip at each interaction & trial summary & float & 0 & 1 \\
$n$ & The number of turns & trial summary & integer & 1 & 200 \\
$p_e$ & The probability of a match ending in the next turn & trial summary & float & 0 & 1 \\
$N$ & The number of strategies in the tournament & trial summary & integer & 3 & 195 \\
$k$ & The number of repetitions of a given tournament & trial summary & integer & 10 & 100 \\
    \bottomrule
        \end{tabular}}
    \end{center}
    \caption{\textbf{Included features for performance evaluation analysis.}
    Stochastic, makes use of length and makes use of game are APL
    classifiers that determine whether a strategy is stochastic or deterministic,
    whether it makes use of the number of turns or the game's payoffs. The
    memory usage is calculated as the number of turns the strategy considers to
    make an action (which is specified in the APL) divided by the number of
    turns. The SSE (introduced in~\cite{Knight2019}) shows how close a strategy
    is to behaving as a ZDs, and subsequently, in an extortionate way. The
    method identifies the ZDs closest to a given strategy and calculates the
    algebraic distance between them as the sum of squared error (SSE). A SSE value of 1 indicates
    no extortionate behaviour at all whereas a value of 0 indicates that a
    strategy is behaving as a ZDs. The
    memory usage of strategies is the number of rounds of play used by the strategy
    when deciding on an action,  divided by the number of turns in each match. For
    example, Winner12 uses the previous two rounds of play, and if participating in
    a match with 100 turns its memory usage would be 2/100. For strategies with an
    infinite memory size, for example Evolved FSM 16 Noise 05, memory usage is equal
    to 1. Note that for tournaments with a probabilistic ending the number of turns
    was not collected, so the memory usage feature is not used for probabilistic
    ending tournaments. The rest of the features considered are the $CC$
    to $C$, $CD$ to $C$, $DC$ to $C$, and $DD$ to $C$ rates as well as
    cooperating ratio of a strategy, the minimum (\(C_{min}\)), maximum
    (\(C_{max}\)), mean (\(C_{mean}\)) and median (\(C_{median}\)) cooperating
    ratios of each tournament.}
    \label{table:manual_features}
\end{table}

\newcolumntype{g}{>{\columncolor{Gray}}c}
\begin{table}[!htbp]
    \begin{center}
    \resizebox{.8\textwidth}{!}{
        \begin{tabular}{lggccggccggg}
    \toprule
    &  \multicolumn{2}{g}{Standard} & \multicolumn{2}{c}{Noisy $p_n \leq 0.1$} & \multicolumn{2}{g}{Probabilistic ending $p_e \leq 0.1$} &  \multicolumn{2}{c}{Noisy probabilistic ending} \\
\midrule
{} &  $r$ &  median score &  $r$ &  median score &  $r$ &  median score &  $r$ &  median score \\
\midrule
$CC$ to $C$ rate & -0.501 & 0.501 & -0.210 & 0.194 & -0.336 & 0.348 & 0.087 & 0.015 \\
$CD$ to $C$ rate & 0.226 & -0.199 & 0.337 & -0.235 & 0.458 & -0.352 & 0.609 & -0.372 \\
$DC$ to $C$ rate & 0.127 & -0.100 & 0.227 & -0.111 & 0.164 & -0.105 & 0.410 & -0.203 \\
$DD$ to $C$ rate & 0.412 & -0.396 & 0.549 & -0.391 & 0.433 & -0.378 & 0.615 & -0.407 \\
$C_r$ & -0.323 & 0.383 & 0.298 & -0.051 & -0.060 & 0.160 & 0.595 & -0.213 \\
$C_{max}$ & 0.000& 0.050 & -0.000& 0.244 & -0.000& 0.079 & -0.000& 0.296 \\
$C_{min}$ & 0.000& 0.085 & 0.000& -0.070 & 0.000& 0.128 & 0.000& 0.000\\
$C_{median}$ & 0.000& 0.209 & 0.000& 0.572 & -0.000& 0.324 & 0.000& 0.667 \\
$C_{mean}$ & 0.000& 0.229 & -0.000& 0.583 & -0.000& 0.354 & -0.000& 0.689 \\
$C_r$ / $C_{max}$  & -0.323 & 0.381 & 0.307 & -0.076 & -0.060 & 0.156 & 0.608 & -0.246 \\
$C_{min}$ / $C_r$ & 0.109 & -0.080 & -0.141 & -0.011 & 0.024 & 0.029 & -0.335 & 0.092 \\
$C_r$ / $C_{median}$ & -0.330 & 0.353 & 0.326 & -0.258 & -0.065 & 0.111 & 0.614 & -0.464 \\
$C_r$ / $C_{mean}$ & -0.331 & 0.357 & 0.325 & -0.228 & -0.066 & 0.114 & 0.617 & -0.431 \\
$N$ & -0.000& -0.009 & -0.000& -0.017 & -0.000& 0.011 & 0.000& 0.139 \\
$k$ & -0.000& -0.002 & -0.000& -0.003 & -0.000& 0.010 & -0.000& 0.035 \\
$n$ & -0.000& -0.125 & -0.000& -0.392 & - & - & - & - \\
$p_n$ & - & - & 0.000& -0.244 & - & - & 0.000& -0.272 \\
$p_e$ & - & - & - & - & 0.000& 0.257 & 0.000& 0.568 \\
Make use of game & -0.003 & -0.022 & -0.047 & 0.014 & -0.046 & 0.022 & -0.110 & 0.057 \\
Make use of length & -0.158 & 0.124 & -0.224 & 0.139 & -0.173 & 0.128 & -0.206 & 0.115 \\
SSE & 0.473 & -0.452 & 0.589 & -0.412 & 0.458 & -0.418 & 0.571 & -0.383 \\
stochastic & 0.006 & -0.024 & 0.010 & -0.007 & -0.001 & 0.001 & -0.001 & 0.002 \\
memory usage & -0.098 & 0.108 & -0.080 & 0.114 & - & - & - & - \\
\bottomrule
\end{tabular}
    }
\end{center}
\caption{\textbf{Correlations between the features of Table~\ref{table:manual_features}
and the normalised rank and the median score.}
The correlation coefficients are calculated using the Spearman's rank
correlation coefficient.}\label{table:correlations}
\end{table}

In standard tournaments, the features \(CC\) to \(C\), \(C_r\), \(C_r /
C_{\text{max}}\), and the cooperating ratio compared to \(C_{\text{median}}\)
and \(C_{\text{mean}}\) have a moderately negative effect on the normalized rank
(a smaller rank is better) and a moderate positive effect on the median score.
The SSE error and the \(DD\) to \(C\) rate have the opposite effects. Thus, in
standard tournaments, behaving cooperatively corresponds to a more successful
performance. Even though being nice generally pays off, that does not hold
against defective strategies. Being more cooperative after a mutual defection,
that is not retaliating, is associated with lesser overall success in terms of
normalized rank.
Compared to standard tournaments, in both noisy and noisy probabilistic ending
tournaments, the higher the rates of cooperation, the lower a strategy's success
and median score. A strategy would not want to cooperate more than both the mean
and median cooperator in such settings. In probabilistic ending tournaments, the
cooperation rate of the winners and its relative comparison to the cooperation
rates of the tournament have no effect. The only features that have an effect
are the \(CD\) to \(C\) rate, which is the tendency of a strategy to forgive,
and the SSE rate, which has a positive effect on the normalized rank.

A multivariate linear regression has been fitted to model the relationship
between the features and the normalized rank. Based on the graphical
representation of the correlation matrices given in the Supplementary
Information, several features are highly correlated and have been removed before
fitting the linear regression model. The features included are given in
Table~\ref{table:linear_regression} alongside their corresponding \(p\) values
in distinct tournaments and their regression coefficients. The \(CD\) to \(C\)
rate has a positively statistically significant effect on the normalized rank
across all tournament types. This suggests that being generous tends to
lower one's performance. In the case of probabilistic ending tournaments, the
coefficient of the \(CD\) to \(C\) rate is the highest, indicating that one
should be more provocative in this setting. Similarly, the SEE error rate has a
positive effect on the normalized rank, suggesting that being extortionate pays
off, especially in noisy tournaments. The measures of cooperation, \(C_r\) and
\(C_r / C_{\text{max}}\), also exhibit a significant effect. In noisy
probabilistic ending tournaments, this effect is positive; however, the
coefficient is very close to zero. In other tournament types, the effect is
negative, indicating that one should aim to be less cooperative than the mean
cooperator of the tournament. However, we cannot interpret the result as
suggesting that a strategy should be as uncooperative as possible.

\newcolumntype{g}{>{\columncolor{Gray}}c}
\begin{table}[!htbp]
    \begin{center}
\resizebox{.8\textwidth}{!}{
    \begin{tabular}{lggccggccgg}
\toprule
& \multicolumn{2}{g}{Standard} & \multicolumn{2}{c}{Noisy $p_n \leq 0.1$} & \multicolumn{2}{g}{Probabilistic ending $p_e \leq 0.1$} & \multicolumn{2}{c}{Noisy probabilistic ending} \\
\midrule
& \multicolumn{2}{g}{\(R\) adjusted: 0.541} & \multicolumn{2}{c}{\(R\) adjusted: 0.373} & \multicolumn{2}{g}{\(R\) adjusted: 0.457} & \multicolumn{2}{c}{\(R\) adjusted: 0.537} \\
{} &  Coefficient &  \(p\)-value &  Coefficient &  \(p\)-value &  Coefficient &  \(p\)-value &  Coefficient &  \(p\)-value \\
\midrule
constant             &  0.695 &  0.000 &  0.560 &  0.000 &  0.627 &  0.000 &  0.345 &  0.005 \\
$CC$ to $C$ rate     & -0.042 &  0.000 & -0.163 &  0.000 & -0.046 &  0.000 &  0.032 &  0.029 \\
$CD$ to $C$ rate     &  0.297 &  0.000 &  0.064 &  0.000 &  0.412 &  0.000 &  0.292 &  0.000 \\
$DC$ to $C$ rate     &  0.198 &  0.000 &  0.142 &  0.000 &  0.193 &  0.000 &  0.193 &  0.000 \\
SSE                  &  0.258 &  0.000 &  0.328 &  0.000 &  0.190 &  0.000 &  0.228 &  0.000 \\
$C_{max}$            & -0.068 &  0.000 & -0.048 &  0.214 & -0.040 &  0.347 & -0.011 &  0.936 \\
$C_{min}$            & -0.161 &  0.000 & -0.029 &  0.367 & -0.049 &  0.017 &  0.008 &  0.912 \\
$C_{mean}$           &  0.117 &  0.000 & -0.133 &  0.000 & -0.159 &  0.000 & -0.468 &  0.000 \\
$C_{min}$ / $C_r$    &  0.057 &  0.000 & -0.006 &  0.322 &  0.054 &  0.000 &  0.034 &  0.099 \\
$C_r$ / $C_{mean}$   & -0.468 &  0.000 & -0.073 &  0.000 & -0.150 &  0.000 &  0.094 &  0.000 \\
$k$                  &  0.000 &  0.325 &  0.000 &  0.965 &  0.000 &  0.079 &  0.000 &  0.065 \\
$n$                  &  0.000 &  0.000 &      - &      - &      - &      - &      - &      - \\
memory usage         & -0.010 &  0.000 & -0.008 &  0.000 &      - &      - &      - &      - \\
$C_r$ / $C_{median}$ &      - &      - &  0.069 &  0.001 & -0.142 &  0.000 &      - &      - \\
$p_n$                &      - &      - & -0.131 &  0.010 &      - &      - & -0.278 &  0.048 \\
$p_e$                &      - &      - &      - &      - & -0.071 &  0.016 &  0.320 &  0.024 \\
\bottomrule
\end{tabular}
}
    \end{center}
    \caption{\textbf{Results of multivariate linear regressions with \(r\) as
    the dependent variable.} \(R\) squared is reported for each model. The $R$
    scores of the fitted models indicate their capability to explain some of the
    variation in the median rank. Most of the features have a statistically
    significant effect on the normalized rank.
    A multivariate linear regression has also be fitted on the median score. The
    coefficients and \(p\) values of the features can be found in
    Supplementary Information. Both approaches lead to similar conclusions.}
    \label{table:linear_regression}
\end{table}

The results presented here suggest that generosity/provocation and a strategy's
cooperation rate, particularly in comparison to the tournament averages, are
significant features. The analysis suggests that strategies should be more
generous in noisy tournaments and less generous in probabilistic ending
tournaments. Moreover, strategies should aim to not cooperate more than the mean
cooperator in their tournaments. We note the analysis is limited as we only
consider a linear relationship between these parameters and the rank. To further
investigate the effects of the parameters discussed in this section, we have
conducted a more detailed analysis in the next section, focusing on the
performances of the winners of the tournaments.

\subsection{Features of top performing strategies}\label{section:winning_features}

\begin{figure}[!htbp]
    \centering
        \centering
        \includegraphics[width=\textwidth]{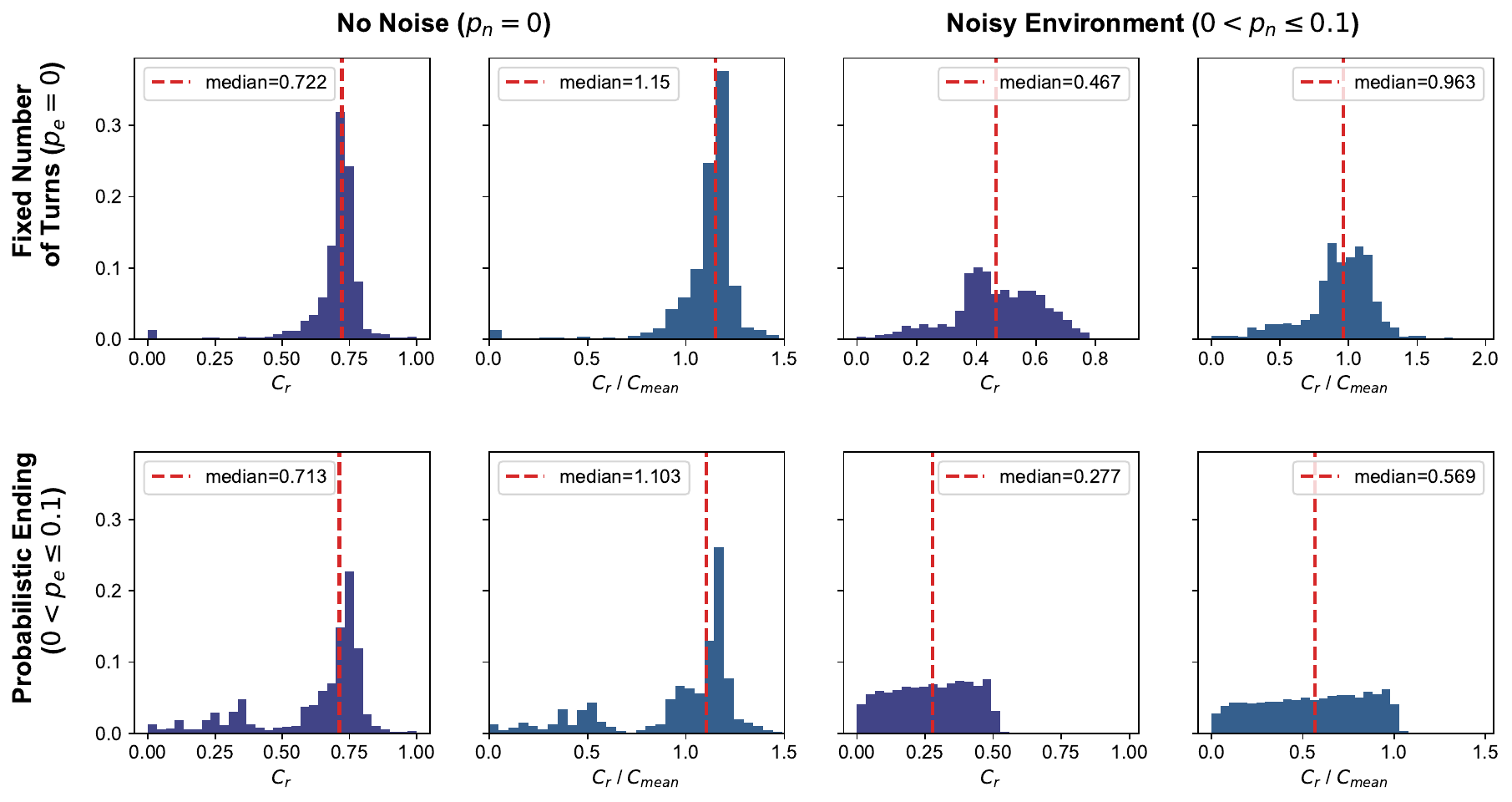}
        \caption{\textbf{Distributions of \(C_r\) and \(C_r / C_{\text{mean}}\)
        for the winners of tournaments.} A value of \(C_r / C_{\text{mean}} =
        1\) imply that the cooperating ratio of the winner was the same as the
        mean cooperating ratio of the tournament.}
        \label{fig:discussion_cooperation_measures}
\end{figure}

In Figure~\ref{fig:discussion_cooperation_measures}, we present the
distributions of the cooperation ratio and \(C_r / C_{\text{mean}}\) for the
winners of tournaments. A value of \(C_r / C_{\text{mean}} = 1\) implies that
the cooperation ratio of the winner was the same as the mean cooperating ratio
of the tournament, and we observe that this occurs for most tournament types,
apart from the case of noisy and probabilistically ending tournaments. In the case of
probabilistic ending tournaments, there are several winners that cooperated much
less than that, confirming the results of the previous section that defecting
strategies can be winners in probabilistic ending tournaments. The distribution
of the cooperation rates showcases a high cooperation rate in standard
tournaments and probabilistic ending tournaments. In tournaments with noise, we
observe a much less cooperative behavior, which could result from strategies
being cautious of potential flip actions by the co-player or strategies not
suited for noise holding grudges against defections.

\begin{figure}[!htbp]
    \centering
        \centering
        \includegraphics[width=.75\textwidth]{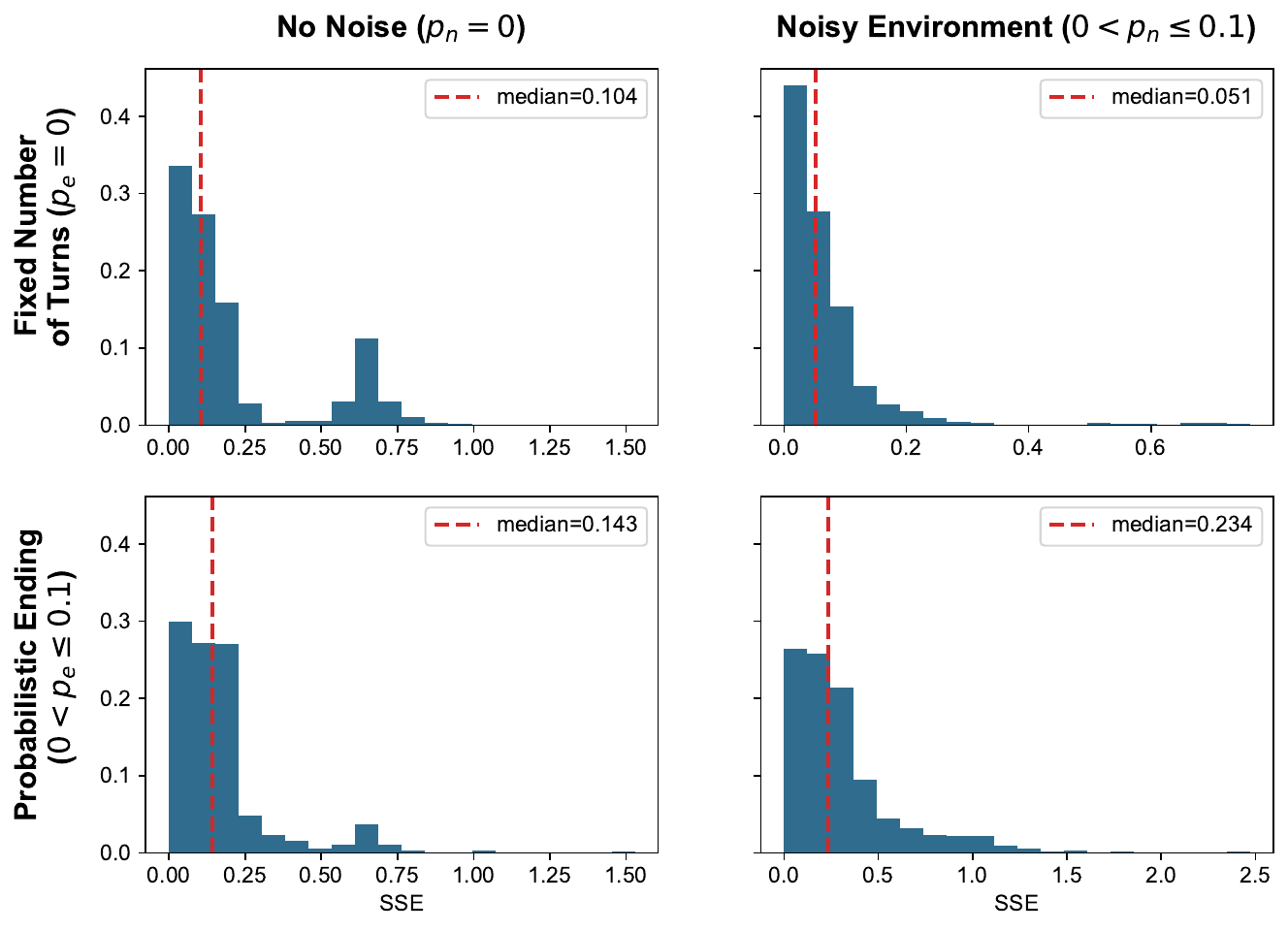}
        \caption{\textbf{Distributions of SSE error for the winners of
        tournaments.} As a reminder, the SSE error shows how close a strategy is
        to behaving as a ZDs, and subsequently, in an extortionate way. A SSE
        value of 1 indicates no extortionate behaviour at all whereas a value of
        0 indicates that a strategy is behaving as a ZDs.}
        \label{fig:discussion_sse}
\end{figure}

Analyzing the SSE distributions across different tournament types
(Figure~\ref{fig:discussion_sse}) suggests that successful strategies exhibit
some extortionate behavior, though not consistently. ZDs are a set of strategies
that are often envious, as they attempt to exploit their opponents. The winners
of the tournaments considered in this work demonstrate envious behavior, but not
to the extent observed in many ZDs. While the exact interactions between matches
are not recorded here, the work of~\cite{Harper2017}, which introduced the
trained strategies appearing in the top-ranked strategies of
Section~\ref{section:top_performances}, did record such interactions.
In~\cite{Harper2017}, it was shown that clever strategies managed to achieve
mutual cooperation with stronger strategies while exploiting weaker ones. This
could explain the clever winners in our analysis and the observed SSE
distributions.

This might also be the reason why ZDs fail to appear in the top ranks—they
attempt to exploit all opponents and cannot actively adapt back to mutual
cooperation against stronger strategies, which requires a deeper memory. It's
worth noting that ZDs tend to perform poorly in population games for a similar
reason: they aim to exploit other players using ZDs, failing to form a
cooperative subpopulation~\cite{Knight2017evolution}. This makes them effective
invaders but poor at resisting invasion.

\begin{figure}[!htbp]
    \centering
        \centering
        \includegraphics[width=\textwidth]{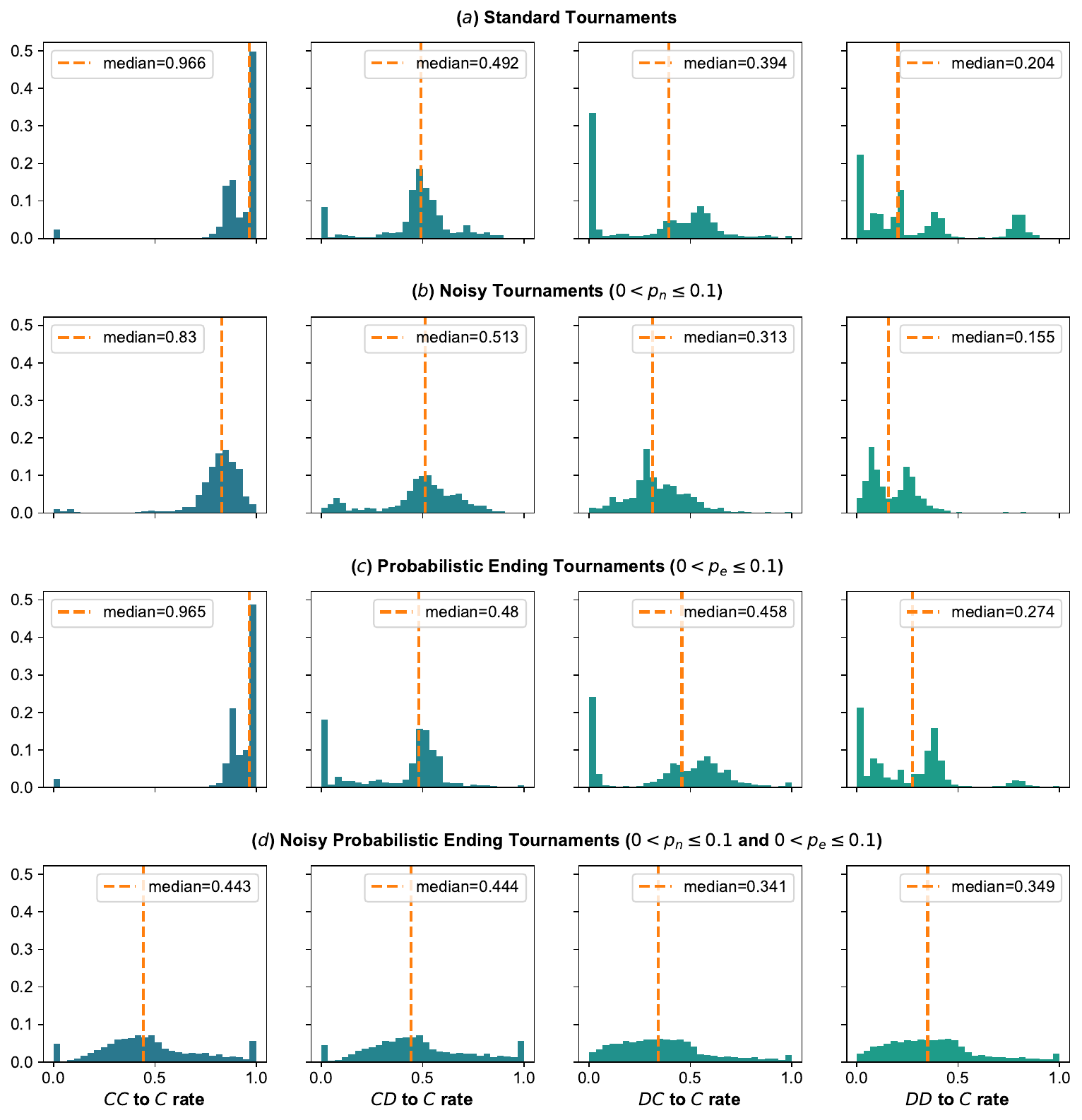}
        \caption{\textbf{Distributions of rates $CC$ to $C$, $CD$ to $C$, $DC$
        to $C$, and $DD$ to $C$ for the winners of
        tournaments.}}\label{fig:discussion_rates}
\end{figure}

Finally, we examine the distributions of the cooperation rates after the
outcomes \(CC, CD, DC\), and \(DD\), as shown in
Figure~\ref{fig:discussion_rates}. In the case of cooperating after mutual
cooperation, the results align with expectations; the distributions skew towards
1, indicating that the winners of the tournaments are more likely to cooperate
after mutual cooperation. Regarding the \(CD\) outcome and the likelihood to
cooperate after such a result, capturing generosity, the distributions skew
towards 1/2, not 1, suggesting that strategies need to reduce their readiness to
forgive. This aligns with the known result that Generous Tit For Tat generally
outperforms \TFT{} in most settings. In probabilistic ending tournaments, there
is a peak at 0, suggesting that strategies should not be too generous in
tournaments with short matches. Such a peak also appears in standard
tournaments; however, not in tournaments with noise, where a strategy should be
more generous.

Part of a strategy's envious behavior can be captured by the rate of \(DC\) to
\(C\). In noisy tournaments, winners are not too envious, but in tournaments
without noise, we can see that winners behave in two ways. Some are a bit
envious, whereas others are very envious. In the \(DD\) to \(D\), we can observe
that, expectedly, the results are skewed towards 0. However, there are winners
that attempt to recover from a \(DD\) outcome. The remaining results are as
expected, skewed towards 0.

\section{Discussion}\label{section:conclusion}

This manuscript explores the performance of 195 strategies in
the \IPD{} in thousands of computer tournaments. The collection of computer
tournaments presented here is the largest and most diverse in the literature.
The 195 strategies are drawn from \axelrod{} library and include
strategies from the \IPD{} literature. The computer tournaments encompass four
different types. So, what is the best way to play the \IPD{}? And is there a
single dominant strategy for the \IPD{}? There was not a single strategy within
the collection of 195 strategies that managed to perform well in
all the tournament variations it competed in. A strategy ranking highly in a
specific environment did not guarantee its success over different tournament
types, with a few exceptions - strategies that generalize better. Already
well-known in the AI/ML literature, adding noise to training data leads to more
robust models. We see that clearly here, where the strategies trained for noise
(or designed for noise) tend to be better generalists. There were instances
where a few strategies trained in narrow conditions outperformed more generalist
strategies, as they tend to overfit. However, the strategies trained with noise
perform well in general, whilst the strategies trained specifically on no noise
or small subpopulations do not.

We also examined the best-performing strategies across various tournament
types and analyzed their salient features. This demonstrated that there are
properties associated with the success of strategies that contradict the
originally suggested properties of Axelrod~\cite{Axelrod1981}.
We showed that complex or \textbf{clever} strategies can be effective, whether
trained against a corpus of possible opponents or purposely designed to mitigate
the impact of noise such as the DBS strategy. Moreover, we found some
strategies designed or trained for noisy environments were also highly ranked in
noise-free tournaments which reinforces the idea that strategies'
complexity/cleverness is not necessarily a liability, rather it can confer
adaptability to a more diverse set of environments.
We also showed that while the type of exploitation attempted by ZDs is
not typically effective in standard tournaments, \textbf{envious} strategies
capable of both exploiting and not their opponents can be highly successful.
Based on the results of~\cite{Harper2017} this could be because they are
selectively exploiting weaker opponents while mutually cooperating with stronger
opponents. Highly noisy or tournaments with short matches also favoured envious
strategies. These environments mitigated the value of being nice. Uncertainty
enables exploitation, reducing the ability of maintaining or enforcing mutual
cooperation, while triggering grudging strategies to switch from typically
cooperating to typically defecting.

The features analysis of the best performing strategies demonstrated that a
strategy should reciprocate, as suggested by Axelrod, but it should relax its
readiness to do so and be more \textbf{generous}. For noisy environments this is
inline with the results of~\cite{Bendor1991, Donninger1986, Molander1985,
Hammerstein1984}, however, we also showed that generosity pays off even in
standard settings, and that in fact the only setting a strategy would want to be
too provocable is when the matches are not long.
Forgiveness as defined by
Axerlod was not explored here. This was mainly because the two round states were
not recorded during the data collection. This could be a topic of future work
that examines the impact of considering more rounds of history.
The features analysis also concluded that
there is a significant importance in \textbf{adapting to the environment}, and
more specifically, to the mean cooperator. 
In most tournament types, the winner of the tournament was also the average
cooperator. Even in tournaments with short matches where defecting behavior
could secure a win, a large number of winners were average cooperators.

This could potentially explain the early success of \TFT. \TFT{} naturally achieves
a cooperation rate near $C_{\text{mean}}$ by virtue of copying its opponent's
last move while also minimizing instances where it is exploited by an opponent
(cooperating while the opponent defects), at least in non-noisy tournaments. It
could also explain why Tit For \(N\) Tats does not fare well for $N > 1$ -- it
fails to achieve the proper cooperation ratio by tolerating too many defections.

Similarly, our results could suggest an explanation regarding the intuitively
unexpected effectiveness of memory-one strategies historically. Given that among
the important features associated with success are the relative cooperation rate
to the population average and the four memory-one probabilities of cooperating
conditional on the previous round of play, these features can be optimized by a
memory-one strategy such as \TFT. Usage of more history becomes valuable when
there are exploitable opponent patterns. This is indicated by the importance of
SSE as a feature, showing that the first-approximation provided by a memory-one
strategy is no longer sufficient.
These results highlight a central idea in evolutionary game theory in this
context: the fitness landscape is a function of the population (where fitness in
this case is tournament performance). While that may seem obvious now, it shows
why historical tournament results on small or arbitrary populations of
strategies have so often failed to produce generalizable results.

Overall, the five properties successful strategies need to have in a \IPD{} competition
based on the analysis that has been presented in this manuscript are:

\begin{enumerate}[(i)]
    \item Be ``nice'' in non-noisy environments or when game lengths are longer
    \item Be provocable in tournaments with short matches, and generous in tournaments with noise
    \item Be a little bit envious
    \item Be clever
    \item Adapt to the environment (including the population of strategies).
\end{enumerate}

The results presented here were based only on a subset of the whole data we have
collected. The analysis of the full dataset is discussed in the Supplementary Information.
However, we can see that the general results of our work remain the same. In the
Supplementary Information, we also evaluate the importance of features using a
random forest classifier and a clustering approach. The results of these
analyses are also in line with the results presented here.

The data set described in this work contains the largest number of \IPD{}
tournaments, to the authors knowledge. The raw data set is available
at~\cite{raw_data} and the processed data at~\cite{data}. Further data mining
could be applied and provide new insights in the field.

\section{Data availability statement}

The raw and processed datasets have been made publicly available~\cite{data,
raw_data} and can be used for further analysis and insights.

\section{Acknowledgements}

N.G. acknowledges the generous support of the European Research Council Starting
Grant 850529: E-DIRECT and the Max Planck Society.
A variety of open-source software have been used in this work. The authors would
like to express their gratitude to the open-source software community, whose
invaluable contributions significantly enhanced the development and execution of
this research. Namely, the authors would like to thank the developers of the
following software packages: Axelrod-Python library for IPD
simulations~\cite{axelrodproject}, the Matplotlib library for
visualisation~\cite{hunter2007matplotlib}, The Numpy library for data
manipulation~\cite{walt2011numpy}, and finally the scikit-learn library for data
analysis~\cite{scikit-learn}.

\bibliographystyle{naturemag}
\bibliography{bibliography}

\begin{thebibliography}{10}
\expandafter\ifx\csname url\endcsname\relax
  \def\url#1{\texttt{#1}}\fi
\expandafter\ifx\csname urlprefix\endcsname\relax\def\urlprefix{URL }\fi
\providecommand{\bibinfo}[2]{#2}
\providecommand{\eprint}[2][]{\url{#2}}

\bibitem{Axelrod1981}
\bibinfo{author}{Axelrod, R.} \& \bibinfo{author}{Hamilton, W.~D.}
\newblock \bibinfo{title}{The evolution of cooperation}.
\newblock \emph{\bibinfo{journal}{science}} \textbf{\bibinfo{volume}{211}},
  \bibinfo{pages}{1390--1396} (\bibinfo{year}{1981}).

\bibitem{Flood1958}
\bibinfo{author}{Flood, M.~M.}
\newblock \bibinfo{title}{Some experimental games}.
\newblock \emph{\bibinfo{journal}{Management Science}}
  \textbf{\bibinfo{volume}{5}}, \bibinfo{pages}{5--26} (\bibinfo{year}{1958}).
\newblock \urlprefix\url{https://doi.org/10.1287/mnsc.5.1.5}.
\newblock \eprint{https://doi.org/10.1287/mnsc.5.1.5}.

\bibitem{Axelrod1980a}
\bibinfo{author}{Axelrod, R.}
\newblock \bibinfo{title}{Effective choice in the prisoner's dilemma}.
\newblock \emph{\bibinfo{journal}{Journal of Conflict Resolution}}
  \textbf{\bibinfo{volume}{24}}, \bibinfo{pages}{3--25} (\bibinfo{year}{1980}).
\newblock \urlprefix\url{https://doi.org/10.1177/002200278002400101}.
\newblock \eprint{https://doi.org/10.1177/002200278002400101}.

\bibitem{Axelrod1980b}
\bibinfo{author}{Axelrod, R.}
\newblock \bibinfo{title}{More effective choice in the prisoner's dilemma}.
\newblock \emph{\bibinfo{journal}{Journal of Conflict Resolution}}
  \textbf{\bibinfo{volume}{24}}, \bibinfo{pages}{379--403}
  (\bibinfo{year}{1980}).

\bibitem{Beaufils1997}
\bibinfo{author}{Beaufils, B.}, \bibinfo{author}{Delahaye, J.~P.} \&
  \bibinfo{author}{Mathieu, P.}
\newblock \bibinfo{title}{Our meeting with gradual, a good strategy for the
  iterated prisoner’s dilemma}.
\newblock In \emph{\bibinfo{booktitle}{Proceedings of the Fifth International
  Workshop on the Synthesis and Simulation of Living Systems}},
  \bibinfo{pages}{202--209} (\bibinfo{year}{1997}).

\bibitem{tzafestas-2000a}
\bibinfo{author}{Tzafestas, E.}
\newblock \bibinfo{title}{Toward adaptive cooperative behavior}
  \textbf{\bibinfo{volume}{2}}, \bibinfo{pages}{334--340}
  (\bibinfo{year}{2000}).

\bibitem{Bendor1991}
\bibinfo{author}{Bendor, J.}, \bibinfo{author}{Kramer, R.~M.} \&
  \bibinfo{author}{Stout, S.}
\newblock \bibinfo{title}{When in doubt... cooperation in a noisy prisoner's
  dilemma}.
\newblock \emph{\bibinfo{journal}{The Journal of Conflict Resolution}}
  \textbf{\bibinfo{volume}{35}}, \bibinfo{pages}{691--719}
  (\bibinfo{year}{1991}).
\newblock \urlprefix\url{http://www.jstor.org/stable/174072}.

\bibitem{Donninger1986}
\bibinfo{author}{Donninger, C.}
\newblock \emph{\bibinfo{title}{Is it Always Efficient to be Nice? A Computer
  Simulation of Axelrod's Computer Tournament}}
  (\bibinfo{publisher}{Physica-Verlag HD}, \bibinfo{address}{Heidelberg},
  \bibinfo{year}{1986}).
\newblock \urlprefix\url{https://doi.org/10.1007/978-3-642-95874-8_9}.

\bibitem{Molander1985}
\bibinfo{author}{Molander, P.}
\newblock \bibinfo{title}{The optimal level of generosity in a selfish,
  uncertain environment}.
\newblock \emph{\bibinfo{journal}{The Journal of Conflict Resolution}}
  \textbf{\bibinfo{volume}{29}}, \bibinfo{pages}{611--618}
  (\bibinfo{year}{1985}).
\newblock \urlprefix\url{http://www.jstor.org/stable/174244}.

\bibitem{Hammerstein1984}
\bibinfo{author}{Selten, R.} \& \bibinfo{author}{Hammerstein, P.}
\newblock \bibinfo{title}{Gaps in harley's argument on evolutionarily stable
  learning rules and in the logic of “tit for tat”}.
\newblock \emph{\bibinfo{journal}{Behavioral and Brain Sciences}}
  \textbf{\bibinfo{volume}{7}}, \bibinfo{pages}{115–116}
  (\bibinfo{year}{1984}).

\bibitem{Nowak1992}
\bibinfo{author}{Nowak, M.~A.} \& \bibinfo{author}{Sigmund, K.}
\newblock \bibinfo{title}{Tit for tat in heterogeneous populations}.
\newblock \emph{\bibinfo{journal}{Nature}} \textbf{\bibinfo{volume}{355}},
  \bibinfo{pages}{250} (\bibinfo{year}{1992}).

\bibitem{Nowak1993}
\bibinfo{author}{Nowak, M.} \& \bibinfo{author}{Sigmund, K.}
\newblock \bibinfo{title}{A strategy of win-stay, lose-shift that outperforms
  tit-for-tat in the prisoner's dilemma game}.
\newblock \emph{\bibinfo{journal}{Nature}} \textbf{\bibinfo{volume}{364}},
  \bibinfo{pages}{56} (\bibinfo{year}{1993}).

\bibitem{kendall2007iterated}
\bibinfo{author}{Kendall, G.}, \bibinfo{author}{Yao, X.} \&
  \bibinfo{author}{Chong, S.~Y.}
\newblock \emph{\bibinfo{title}{The iterated prisoners' dilemma: 20 years on}},
  vol.~\bibinfo{volume}{4} (\bibinfo{publisher}{World Scientific},
  \bibinfo{year}{2007}).

\bibitem{Press2012}
\bibinfo{author}{Press, W.~H.} \& \bibinfo{author}{Dyson, F.~J.}
\newblock \bibinfo{title}{Iterated prisoner’s dilemma contains strategies
  that dominate any evolutionary opponent}.
\newblock \emph{\bibinfo{journal}{Proceedings of the National Academy of
  Sciences}} \textbf{\bibinfo{volume}{109}}, \bibinfo{pages}{10409--10413}
  (\bibinfo{year}{2012}).

\bibitem{Stewart2012}
\bibinfo{author}{Stewart, A.~J.} \& \bibinfo{author}{Plotkin, J.~B.}
\newblock \bibinfo{title}{Extortion and cooperation in the prisoner’s
  dilemma}.
\newblock \emph{\bibinfo{journal}{Proceedings of the National Academy of
  Sciences}} \textbf{\bibinfo{volume}{109}}, \bibinfo{pages}{10134--10135}
  (\bibinfo{year}{2012}).

\bibitem{mathieu2017}
\bibinfo{author}{Mathieu, P.} \& \bibinfo{author}{Delahaye, J.~P.}
\newblock \bibinfo{title}{New winning strategies for the iterated prisoner's
  dilemma}.
\newblock \emph{\bibinfo{journal}{Journal of Artificial Societies and Social
  Simulation}} \textbf{\bibinfo{volume}{20}}, \bibinfo{pages}{12}
  (\bibinfo{year}{2017}).
\newblock \urlprefix\url{http://jasss.soc.surrey.ac.uk/20/4/12.html}.

\bibitem{Harper2017}
\bibinfo{author}{Harper, M.} \emph{et~al.}
\newblock \bibinfo{title}{Reinforcement learning produces dominant strategies
  for the iterated prisoner’s dilemma}.
\newblock \emph{\bibinfo{journal}{PloS one}} \textbf{\bibinfo{volume}{12}},
  \bibinfo{pages}{e0188046} (\bibinfo{year}{2017}).

\bibitem{Axelrod1987}
\bibinfo{author}{Axelrod, R.}
\newblock \bibinfo{title}{The evolution of strategies in the iterated
  prisoner's dilemma}.
\newblock \emph{\bibinfo{journal}{Genetic Algorithms and Simulated Annealing}}
  \bibinfo{pages}{32--41} (\bibinfo{year}{1987}).
\newblock \urlprefix\url{http://ci.nii.ac.jp/naid/10000082922/en/}.

\bibitem{Miller1996}
\bibinfo{author}{Miller, J.~H.}
\newblock \bibinfo{title}{The coevolution of automata in the repeated
  prisoner's dilemma}.
\newblock \emph{\bibinfo{journal}{Journal of Economic Behavior and
  Organization}} \textbf{\bibinfo{volume}{29}}, \bibinfo{pages}{87 -- 112}
  (\bibinfo{year}{1996}).
\newblock
  \urlprefix\url{http://www.sciencedirect.com/science/article/pii/0167268195000526}.

\bibitem{axelrodproject}
\bibinfo{author}{project developers, T.~A.}
\newblock \bibinfo{title}{Axelrod: 3.0.0}.
\newblock \bibinfo{howpublished}{\url{http://dx.doi.org/10.5281/zenodo.807699}}
  (\bibinfo{year}{2016}).

\bibitem{Au2006}
\bibinfo{author}{Au, T.~C.} \& \bibinfo{author}{Nau, D.}
\newblock \bibinfo{title}{Accident or intention: that is the question (in the
  noisy iterated prisoner's dilemma)}.
\newblock In \emph{\bibinfo{booktitle}{Proceedings of the fifth international
  joint conference on Autonomous agents and multiagent systems}},
  \bibinfo{pages}{561--568} (\bibinfo{organization}{ACM},
  \bibinfo{year}{2006}).

\bibitem{prison}
\bibinfo{title}{Lifl (1998) prison}.
\newblock \bibinfo{howpublished}{\url{http://www.lifl.fr/IPD/ipd.frame.html}}.
\newblock \bibinfo{note}{Accessed: 2017-10-23}.

\bibitem{Eckhart2015}
\bibinfo{author}{A., E.}
\newblock \bibinfo{title}{Coopsim v0.9.9 beta 6}.
\newblock \bibinfo{howpublished}{\url{https://github.com/jecki/CoopSim/}}
  (\bibinfo{year}{2015}).

\bibitem{lesswrong}
\bibinfo{author}{prase}.
\newblock \bibinfo{title}{Prisoner's dilemma tournament results}.
\newblock
  \bibinfo{howpublished}{\url{https://www.lesswrong.com/posts/hamma4XgeNrsvAJv5/prisoner-s-dilemma-tournament-results}}
  (\bibinfo{year}{2011}).

\bibitem{Ashlock2006}
\bibinfo{author}{Ashlock, W.} \& \bibinfo{author}{Ashlock, D.}
\newblock \bibinfo{title}{Changes in prisoner's dilemma strategies over
  evolutionary time with different population sizes}.
\newblock In \emph{\bibinfo{booktitle}{2006 IEEE International Conference on
  Evolutionary Computation}}, \bibinfo{pages}{297--304}
  (\bibinfo{organization}{IEEE}, \bibinfo{year}{2006}).

\bibitem{Ashlock2014}
\bibinfo{author}{Ashlock, W.}, \bibinfo{author}{Tsang, J.} \&
  \bibinfo{author}{Ashlock, D.}
\newblock \bibinfo{title}{The evolution of exploitation}.
\newblock In \emph{\bibinfo{booktitle}{2014 IEEE Symposium on Foundations of
  Computational Intelligence (FOCI)}}, \bibinfo{pages}{135--142}
  (\bibinfo{organization}{IEEE}, \bibinfo{year}{2014}).

\bibitem{Knight2019}
\bibinfo{author}{Knight, V.~A.}, \bibinfo{author}{Harper, M.},
  \bibinfo{author}{Glynatsi, N.~E.} \& \bibinfo{author}{Gillard, J.}
\newblock \bibinfo{title}{Recognising and evaluating the effectiveness of
  extortion in the iterated prisoner's dilemma}.
\newblock \emph{\bibinfo{journal}{CoRR}}
  \textbf{\bibinfo{volume}{abs/1904.00973}} (\bibinfo{year}{2019}).
\newblock \urlprefix\url{http://arxiv.org/abs/1904.00973}.
\newblock \eprint{1904.00973}.

\bibitem{Knight2017evolution}
\bibinfo{author}{Knight, V.}, \bibinfo{author}{Harper, M.},
  \bibinfo{author}{Glynatsi, N.~E.} \& \bibinfo{author}{Campbell, O.}
\newblock \bibinfo{title}{Evolution reinforces cooperation with the emergence
  of self-recognition mechanisms: an empirical study of the moran process for
  the iterated prisoner's dilemma}.
\newblock \emph{\bibinfo{journal}{arXiv preprint arXiv:1707.06920}}
  (\bibinfo{year}{2017}).

\bibitem{raw_data}
\bibinfo{author}{Glynatsi, N.~E.}
\newblock \bibinfo{title}{{A data set of 45686 Iterated Prisoner's Dilemma
  tournaments' results [RAW DATA]}} (\bibinfo{year}{2023}).
\newblock \urlprefix\url{https://doi.org/10.5281/zenodo.10246248}.

\bibitem{data}
\bibinfo{author}{E., N.}
\newblock \bibinfo{title}{{A data set of 45686 Iterated Prisoner's Dilemma
  tournaments' results}} (\bibinfo{year}{2023}).
\newblock \urlprefix\url{https://doi.org/10.5281/zenodo.10246247}.

\bibitem{hunter2007matplotlib}
\bibinfo{author}{Hunter, J.~D.}
\newblock \bibinfo{title}{{Matplotlib: A 2D graphics environment}}.
\newblock \emph{\bibinfo{journal}{Computing In Science \& Engineering}}
  \textbf{\bibinfo{volume}{9}}, \bibinfo{pages}{90--95} (\bibinfo{year}{2007}).

\bibitem{walt2011numpy}
\bibinfo{author}{Walt, S.}, \bibinfo{author}{Colbert, S.~C.} \&
  \bibinfo{author}{Varoquaux, G.}
\newblock \bibinfo{title}{{The NumPy array: a structure for efficient numerical
  computation}}.
\newblock \emph{\bibinfo{journal}{Computing in Science \& Engineering}}
  \textbf{\bibinfo{volume}{13}}, \bibinfo{pages}{22--30}
  (\bibinfo{year}{2011}).

\bibitem{scikit-learn}
\bibinfo{author}{Pedregosa, F.} \emph{et~al.}
\newblock \bibinfo{title}{Scikit-learn: Machine learning in {P}ython}.
\newblock \emph{\bibinfo{journal}{Journal of Machine Learning Research}}
  \textbf{\bibinfo{volume}{12}}, \bibinfo{pages}{2825--2830}
  (\bibinfo{year}{2011}).

\end{thebibliography}

\includepdf[pages=-]{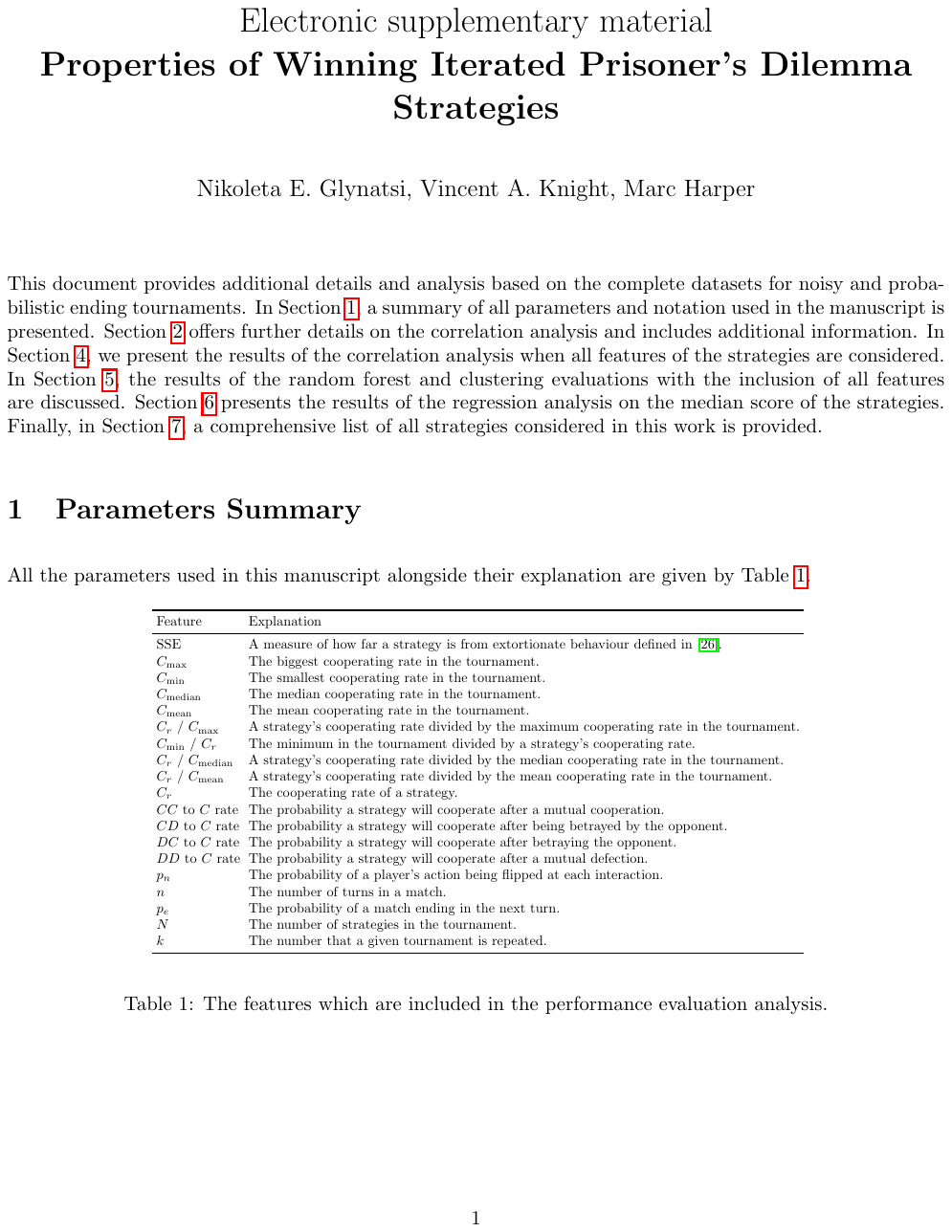}

\end{document}